\documentclass[ reprint,  amsmath,amssymb, aps ]{revtex4-1}

\usepackage{amsmath,amssymb,graphicx}
\usepackage{dcolumn}% Align table columns on decimal point
\usepackage{bm}% bold math
\usepackage{mathtools}
\DeclarePairedDelimiter{\ceil}{\lceil}{\rceil}
\usepackage{natbib}
\usepackage{xcolor}
\begin{document}
	
	\preprint{APS/123-QED}
	
	\title{Tripling the 99 \% mode conversion bandwidth of  gratings using a single phase shift}% Force line breaks with \\
	%\thanks{A footnote to the article title}%
	
	\author{Soham Basu}
	%\altaffiliation{Alumni, School of Engineering, École Polytechnique Fédérale de Lausanne}%Lines break automatically or can be forced with \\
	\email{sohambasu6817@gmail.com}
	
	\date{\today}% It is always \today, today,
	%  but any date may be explicitly specified

\begin{abstract}
	Achieving large bandwidth at strong mode conversion strengths is a central problem in few-mode optical waveguides. \textcolor{black}{The most uncomplicated modification of a standard mode-converting grating is to incorporate a single extra gap. Such simple design has not been applied yet for broadening conversion spectrum between core modes of any few-mode waveguide e.g. few-mode fiber, possibly due to lacking knowledge of critical parameter control:}
	%Externally-written gratings for broadband operation have missed multiple points as yet:  a) use of a single phase shift in a standard grating, 
	%requisite control of the critical 
	\textcolor{black}{(a)} phase shift value at resonance wavelength, and \textcolor{black}{(b)} deterministic method for placement of the phase shift(s) accommodating the change in waveguide properties during grating writing. From semi-analytical considerations, we show \textcolor{black}{in full generality} that our design of a single-phase-shifted grating increases the bandwidth at 99 \% conversion strength by more than 2.9 times\textcolor{black}{, together with elucidating} crucial parameter tolerances. The phase-shift placement is determinable on-the-go from the conversion spectra, thereby obviating any estimation of modified waveguide properties during grating-writing. This offers the longed on-demand in-fiber solution for dispersion control in fiber laser cavities and communication systems.
\end{abstract}

%\keywords{Suggested keywords}%Use showkeys class option if keyword
%display desired
\maketitle

%\tableofcontents

Two-dimensional waveguides support finite number of diffraction-less propagating modes, thereby simplifying the design of photonic devices. Ignoring polarization, the electric field of an LP mode \cite{gloge} is given by 
\begin{eqnarray}
\begin{cases}
E_{\beta_k} (x,y, \lambda) e^{i ( \textcolor{black}{\frac{2 \pi c}{\lambda}} t - \beta_k(\lambda) z)} & \textrm{propagating forward} \\
E_{\beta_k} (x,y, \lambda) e^{i ( \textcolor{black}{\frac{2 \pi c}{\lambda}} t + \beta_k(\lambda) z)} & \textrm{propagating backward} \nonumber
\end{cases}
\end{eqnarray}
where each mode is identified by its unique propagation constant $\beta_k(\lambda)>0$, with transverse electric field $E_{\beta_k}(x,y,\lambda) ~\in \mathbb{R}$ and waveguide axis $z$. The number of modes can be controlled by designing the transverse profile of the waveguide. Operating at single-mode regimes reduce complexity. The enhanced design flexibility of few-mode waveguides come at the expense of increased complexity of differently paced co-propagating modes. Having majority of the power carried by individual modes at different parts the waveguide along with high-purity mode conversion at the intersections circumvents such complexity. A powerful application is dispersion compensation utilizing opposite dispersion signs of different modes \cite{poole_optical_1994}. Near-zero net cavity  dispersion is sought for reducing noise in ultrashort-pulsed fiber laser \cite{namiki_noise_1997}.

Grating mode converters, which are periodic perturbations along the waveguide axis, allow high-purity conversion between co-propagating modes in an inherently mode-selective manner. Power loss from a grating can be made small by keeping the perturbations small and smooth along $z$, at the expense of having longer length.

Standard gratings are limited in bandwidth at strong conversion strengths (Equation \ref{eqn:single_gratings}), often falling short of wavelength ranges of interest e.g. bandwidths of EDFA amplifier for fiber communications and gain of ultrafast fiber lasers. Increasing grating bandwidth has been extensively researched, for example: (a) The seminal solution of achieving mode conversion at flat parts of difference in propagation constants near a turning point wavelength \cite{ramachandran_bandwidth_2002}, limited to the few critical wavelengths, (b) increasing the coupling constant \cite{zheng_all_2016} at the expense of increased loss, especially for gratings asymmetric to the transverse waveguide profile, (c) sophisticated in-fiber phase plates fabricated with laser irradiation \cite{savolainen_broadband_2014} and offset splicing \cite{jung_compact_2015}, and (d) the method of inter-playing between the phases of consequent gratings and  precise gaps interleaving those \cite{chan_analysis_2005}. The ease of fabrication, low-loss and wide wavelength tunability of phase-shifted gratings have attracted wide attention including achievement of \textcolor{black}{broadband} $>99$ \%  mode conversion in bottom-up fabricated polymer waveguide and gratings using an excess of two phase-shifts \cite{wang_ultra-broadband_2017}. Recently,the minimal case of a single phase-shift has been explored in \textcolor{black}{two-mode} optical fibers with dissimilar gratings on the two sides of the phase shift \cite{liu_bandwidth_2020}, which suffer from sub-optimal interplay of phases between the segments. \textcolor{black}{The use of single-phase shifted gratings to broaden 99 \% core-cladding conversion in a single-mode fiber was numerically  simulated and demonstrated with arc-writing \cite{abrishamian_cascade_2015}, showing 3-fold enhancement. Application of such compact design for converting between core modes of waveguides has been surprisingly overlooked. }
%, even though they offer the smallest device footprint and thereby robustness to fiber non-uniformities. }
%We present semi-analytical theoretical proof guaranteeing  performance of single-phase-shifted grating for both core-cladding and core-core mode conversion for any mode pair},
%We show that a standard grating with a single \textcolor{black}{incorporated} phase shift suffices, 
%together with highlighting the criticality of the $\pi$-phase shift at the resonance wavelength.
%Similar performance has yet not been achieved for externally-written grating in pre-made waveguides (e.g. optical fibers), due to lack of understanding of the interplay of parameters which have been altered during the writing process. We present a robust fabrication recipe to increase grating bandwidth by $>2.9$ times compared to a standard grating using a single phase-shift, along with elucidating parameter tolerances. Our simple recipe removes the need to estimate wavelength-dependent parameter changes after the writing process.

Our \textcolor{black}{generalized} fabrication recipe \textcolor{black}{for any mode pair, which is also independent of wavelength-dependent parameter changes during  grating writing, } increases \textcolor{black}{99 \% conversion} bandwidth by $>2.9$ times compared to a standard grating. \textcolor{black}{Semi-analytical proof is presented, along with numerical evidence of the criticality of the $\pi$-phase shift at the resonance wavelength for a particular example.} 

\begin{figure}[ht!]
	\centering
	\includegraphics[trim= 150 220 150 220, clip, width=0.8\columnwidth]{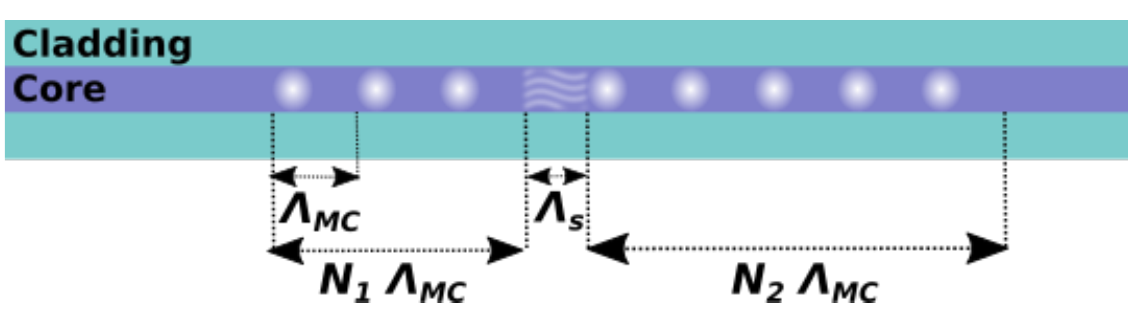}
	\caption{\label{fig:2segs_schematic} Schematic of single-phase-shifted grating of pitch $\Lambda_{MC}$, with $N_1$ and $N_2$ full periods in the two segments with an extra gap of $\Lambda_s$ in between.}
\end{figure}

For a standard grating of pitch $\Lambda_{MC}$ which can achieve $>99 \%$ mode conversion at resonance wavelength $\lambda_{MC}$, let us designate the number of periods corresponding to maximum mode conversion as  $N_0$, which gives the value of coupling per period $K(\lambda_{MC}) \approx \frac{\pi}{2 N_0}$. Compared to the standard grating, greater than $2.9$ times increase the $99$ \% mode conversion bandwidth can be achieved by the following design (figure \ref{fig:2segs_schematic}): (a) first writing $N_1$ periods such that the mode conversion at resonance wavelength $\lambda_{MC}$ just reaches 50 \% or above, (b) followed by an extra gap adding a phase of $\pi$ at $\lambda_{MC}$, (c) followed by writing the minimum $N_2$ number of marks such that $cos^2 \big( K(\lambda_{MC}) (N_2-N_1) \big) \le 10^{-2}$. Additionally, harnessing the simplicity of the linear algebraic model of a single-phase-shifted grating, we identify the bimodal spectral shape of the $>99$ \% mode conversion range.

Perturbations locally change $\beta_k(\lambda) \rightarrow \beta_k(\lambda,z)$. The resonance wavelength $\lambda_{MC}$ is determined by phase-matching over any period length $\Lambda_{MC}$ of the grating $\Phi(\lambda) = \int_z^{z+\Lambda_{MC}} \big( \beta_1(\lambda_{MC}, z)-\beta_2(\lambda_{MC}, z) \big) dz =2 \pi$. Direct method to quantify the additional phase \textcolor{black}{$\Delta \phi(\lambda)$} added per perturbation has recently become available \cite{alcusa-2016, basu}.
\begin{align} \label{eqn:phase_matching}
\Phi(\lambda) = \Lambda_{MC} \big( \beta_1(\lambda_{MC})-\beta_2(\lambda_{MC}) \big) + \Delta \phi (\lambda) \nonumber \\
\end{align}
Using $\frac{\pi \Lambda_{MC}}{\Phi(\lambda_{MC})}$ length of the pristine waveguide to induce $\pi$ phase shift at $\lambda_{MC}$ is erroneous and leads to subpar performance, which may have bottlenecked experimental efforts so-far. Equation \ref{eqn:phase_matching} can additionally be used to estimate both $\lambda_{MC}$ and $\big( \beta_1(\lambda_{MC})-\beta_2(\lambda_{MC}) \big)$ in any new waveguide and writing setup using a single experiment, although that is not a requisite for this method.

%$\textbf{\qquad 2.}$ The maximum mode conversion strength of a standard grating needs to be below 99 \% in order to achieve spectral symmetry for a single $\pi$-phase shifted grating. This is dependent on the ratio of the effective perturbation length to the period which is physically controllable. As for the number and placement of phase shifts, the simplest case of a single phase shift have curiously been ignored. We  show that a single phase shift guarantees $>$2.7 times bandwidth increase at 99 \% conversion strength compared to a standard grating. For bottom-up fabricated waveguides and gratings, the optimal placement of the phase shift can be calculated simulated explicitly. For perturbatively written gratings, it suffices to first write just enough number of periods until 50 \% conversion is achieved at $\lambda_{MC}$, then add a $\pi$-phase shift and finally continue writing until the conversion strength at $\lamba_{MC}$ just crosses $10^{-2}$.

%, which finds uses in dispersion compensation or having a more suitable mode profile at the output.

%Let us designate the number of periods leading to maximum mode conversion for a standard grating as $N_0$. For low-loss gratings, writing $\ceil{\frac{N_0}{2}}$ and $\ceil{2.93 \frac{N_0}{2}}$ number of periods before and after the phase-shift suffices, whereas for high loss more number of periods are needed after the phase shift by this algorithm. 
We will prove our method only for low-loss gratings. In case of non-negligible loss simulated with separate "Discretized-medium" approximation \cite{wang_layer_2001}, we always saw broadband performance for some $N_2 > \ceil{\frac{2.93 N_0}{2}}$ . Power loss per period manifests as both linear shift in $\lambda_{MC}$ during grating-writing and linear decrease in transmission at wavelengths away from the mode conversion footprint. Misalignment of the axes of the waveguide and the translation stage of the grating-writing setup manifests at shifting $\lambda_{MC}$ shifts but negligible loss at faraway wavelengths. 
For low-loss and duty cycle in the range $\big( 0.35, 0.65 \big)$, the complex transmission $M(\lambda,N,\Phi(\lambda), K(\lambda))$ of $N$ grating periods is described by a simple linear algebraic model consisting of the phase $\Phi(\lambda)$ and coupling $K(\lambda)$ per grating period \cite{wang_layer_2001}. Using $\phi(\lambda)=\frac{1}{2} \Phi(\lambda) - \pi$ and $\Gamma(\lambda)= \sqrt{K^2(\lambda) + \phi^2(\lambda)}$ gives $M(\lambda,N,\Phi(\lambda), K(\lambda))=$
\begin{eqnarray} \label{eqn:complex_matrix}
\left[
\begin{array}{cc}
cos(N \Gamma)+i \frac{\phi}{\Gamma} sin(N \Gamma) & i \frac{K}{\Gamma} sin(N \Gamma)\\
i \frac{K}{\Gamma} sin(N \Gamma) & cos(N \Gamma)-i \frac{\phi}{\Gamma}sin(N \Gamma)
\end{array}
\right]  \nonumber \\
\end{eqnarray}
%and for a grating starting at $z$
%\begin{eqnarray}\label{eqn:transfer_matrix_single}
%&& \left[
%\begin{array}{c}
%E_1(\lambda, z +  N \lambda_{MC} ) \\
%E_2(\lambda, z + N \lambda_{MC} ) \\
%\end{array}
%\right] = M(\lambda,N,...) 
%\left[
%\begin{array}{c}
%E_1(\lambda, z ) \\
%E_2(\lambda, z ) \\
%\end{array}
%\right]  
%\end{eqnarray}

%\begin{figure}[ht!]
%    \centering
%    \includegraphics[trim= 0 0 0 0, clip, width=1\columnwidth]{MCGrating}
%    \caption{Grating mode converter}
%    \label{fig:universe}
%\end{figure}
%The complex amplitudes $E_1(\lambda)$ and $E_2(\lambda)$ of two modes interacting with power conservation, over $K$ segments of individual periods $N_1, N_2, ...,N_K$ seperated by $N-1$ phase shifts  described by a 

An extra gap of $\Lambda_s$ after any segment of full periods adds a phase of \textcolor{black}{$\xi(\lambda)=\big( \beta_1(\lambda) - \beta_2(\lambda) \big) \Lambda_s$ without mode conversion, corresponding to $K(\lambda)=0, ~N=1, ~\Phi(\lambda)=\xi(\lambda)$ in equation \ref{eqn:complex_matrix}.}
%An extra gap of $\Lambda_s$ after any segment of full periods adds a phase of $\phi_s(\lambda)=\big( \beta_1(\lambda) - \beta_2(\lambda) \big) \Lambda_s$, which multiplies the complex amplitude by  $P(\lambda,\Lambda_s, \beta_1(\lambda)-\beta_2(\lambda)) 
%=$
%\begin{eqnarray} \label{eqn:phase_matrix}
%\left[
%\begin{array}{cc}
%e^{i\phi_s(\lambda)/2}  & 0 \\
%0 & e^{- i \phi_s(\lambda)/2}
%\end{array}
%\right]  \nonumber \\
%\end{eqnarray}
For $K$ segments with $ N_{1 \le k \le K}$ periods and gaps of $\Lambda_s$ between segments, the \textcolor{black}{complex} transmission \textcolor{black}{at $z_{ \{ N_k \} }=z + \sum_{k=1}^{K} N_k  \textcolor{black}{\Lambda_{MC}} + K \Lambda_{s} $} is given by	

\begin{align}\label{eqn:transfer_matrix}
\textstyle
& \left[
\begin{array}{c}
E_1(\lambda, \textcolor{black}{z_{ \{ N_k \} }} ) \\
E_2(\lambda,  \textcolor{black}{z_{ \{ N_k \} }} ) \\
\end{array}
\right] = T(\lambda, \textcolor{black}{ \{ N_k \}}) 
\left[
\begin{array}{c}
E_1(\lambda, z ) \\
E_2(\lambda, z ) \\
\end{array}
\right]  \nonumber \\
& ~ T(\lambda, \{ N_k \}) =  
\Pi_{k=1}^{K} \Big( \textcolor{black}{~ M(\lambda, \{ 1 \}, \xi(\lambda), 0)} \nonumber \\
& \qquad \qquad \qquad \qquad ~~~ * M(\lambda, \textcolor{black}{\{} N_k \textcolor{black}{\}},\Phi \textcolor{black}{(\lambda)}, K\textcolor{black}{(\lambda)}) ~ \Big)
\nonumber \\
\end{align}

For \textcolor{black}{$K=2$} with an extra gap introducing a phase shift of $\xi(\lambda)$, the entries of the \textcolor{black}{complex} transmission matrix $T_{i,j}(\lambda, N_2, N_1), \quad i, j= 1,2 $ in equation \ref{eqn:transfer_matrix} are especially simple to write 
%using the abbreviations $f(N) = cos(\Gamma(\lambda) N ) + i \frac{\phi(\lambda)}{\Gamma(\lambda)} sin( \Gamma(\lambda) N )$ and $g(N) = \frac{K(\lambda)}{\Gamma(\lambda)} sin( \Gamma(\lambda) N )$
. Since for low-loss gratings $\vert T_{1,1} \vert^2 + \vert T_{2,1} \vert^2 = \vert T_{1,2} \vert^2 + \vert T_{2,2} \vert^2 \approx 1$ and $T_{2,2} = \overline{T_{1,1}}$, only studying the spectral response of $\vert T_{1,1} \vert^2$ suffices.
%, where
%\begin{align}
%& T_{1,1}(\lambda, \{ N_1, N_2 \} ) = e^{i \xi(\lambda)} %f(N_2) f(N_1) - e^{-i \xi(\lambda)} g(N_2) g(N_1) \nonumber \\
%%T_{2,1}(\lambda, N_1, N_2) =& i \Big( e^{i \xi(\lambda)} f(N_2) g(N_1) + e^{-i \xi(\lambda)} g(N_2) \overline{f}(N_2)  \Big) \nonumber \\
%%T_{1,2}(\lambda, N_1, N_2) =& i \Big( e^{i \xi(\lambda)} g({N_2}) f({N_1}) + e^{-i \xi(\lambda)}  \overline{f}({N_2}) g({N_1})  \Big) \nonumber \\
%%T_{2,2}(\lambda, N_1, N_2) =& e^{-i \xi(\lambda)} \overline{f}({N_2}) \overline{f}({N_1}) - e^{i \xi(\lambda)} g({N_2}) g({N_1}) \nonumber
%\end{align}

For \textcolor{black}{conciseness}, let us use the notations $\alpha_1=(N_2 - N_ 1)/N_0$, $\alpha_2=(N_2 + N_ 1)/N_0$, $K=K(\lambda)$, $\phi=\phi(\lambda)$, $\xi=\xi(\lambda)$ and $C(\lambda)=\Gamma(\lambda) N_0$.  Using the normalized parameter ${x \textcolor{black}{ (\lambda)= \frac{\Gamma(\lambda)}{K(\lambda)} = }( 1+\frac{\phi^2(\lambda)}{K^2(\lambda)} )^{\frac{1}{2}}} \ge 1$ we get

\begin{align} 
&e^{-i \xi(\lambda)} T_{1,1}(\lambda, \{ N_1, N_2 \} ) = \nonumber \\ 
& \textcolor{black}{\frac{K^2}{\Gamma^2}} cos(2\xi) cos( (N_2  -N_1) \Gamma ) + \textcolor{black}{\frac{\phi^2}{\Gamma^2}} cos((N_2 + N_1) \Gamma  ) \nonumber \\
&+ i \Bigg( \textcolor{black}{\frac{\phi}{\Gamma}} sin(\Gamma (N_2-N_1) )  
+ \textcolor{black}{\frac{K^2}{\Gamma^2}} sin(2 \xi) sin(N_2\Gamma ) sin(N_1\Gamma) \Bigg) \nonumber \nonumber \\
&= \frac{1}{x^2} cos(2\xi) cos(\alpha_1 C x)  +  ( 1 - \frac{1}{x^2} ) cos( \alpha_2 C x) + \nonumber \\
& i \Bigg( \textcolor{black}{\sqrt{1-\frac{1}{x^2}}} sin(\alpha_2 C x)  
+ \frac{1}{x^2} sin(2 \xi) sin(\frac{N_2}{N_0} C x) sin(\frac{N_1}{N_0} C x) \Bigg) 
\label{eqn:normalized_tr_two_segments} \nonumber \\
\end{align}

For a standard grating with maximum mode conversion ($N_1=N_0, ~N_2=0$), $\vert T_{1,1} \vert^2 \le 10^{2} $ is satisfied for 
\begin{align} \label{eqn:single_gratings}
1 \le x(\lambda) \le 1.005 \nonumber \\
\end{align} 	

With $\xi(x(\lambda_{MC})=1) = \pi$,  we do some simplifications:
\begin{itemize}
	\item Assume $C(\lambda) = K(\lambda) N_0 \approx K(\lambda_{MC}) N_0 = \frac{\pi}{2}$ to be constant based on $\frac{\partial K(\lambda)}{ \partial \lambda} << \frac{\partial \phi(\lambda)}{ \partial \lambda}$ \cite{wang_ultra-broadband_2017}.
	\item Since $\xi(x)$ is unknown without further experimentation \cite{basu}, use bounded variation $g(x,-s) \le 2 \xi(x) \le g(x,s) = 2 \pi (1+ sx-s) $ to normalize $sin(2 \xi (\lambda))$. For most waveguides, $s=0.02$ suffices.
	\item Replace $cos(2 \xi(x(\lambda)))$ by 1, since $cos(2 \xi(x(\lambda)))  \approx 1.0000$ for any $1 \le x(\lambda) \le 1.045$ and $0 \le s \le 0.02$. 
\end{itemize}

With these simplifications, finding the range $1 \le x(\lambda) \le x_{0.01}(N_0,s)$ satisfying $\vert T_{1,1} \vert^2 \le \textcolor{black}{0.01}$ is equivalent to finding $x_{\textcolor{black}{0.01}}(N_0,s) = \min_{-s \le s^* \le s} x_{0.01} (N_0, s^*)$ such that $u^2(x_{0.01}(N_0, s^*), N_0, ~\textcolor{black}{s^*})+v^2(x_{0.01}(N_0, s^*), N_0,s^*) = \textcolor{black}{0.01}$. The material independent $v(x, N_0,s^*)$ and $u(x,N_0, s^*)$ are given by equations \ref{eqn:simplified_v} and \ref{eqn:simplified_u}.

\begin{align} 
&v(x, N_0, ~\textcolor{black}{s^*})=\textcolor{black}{\sqrt{1-\frac{1}{x^2}}} ~ sin(\alpha_2 \pi x) +  \nonumber \\
&  \quad \bigg[ \frac{1}{x^2} sin \bigg( g(x,s^*) sin(\frac{\alpha_1+\alpha_2}{2} \pi x) sin(\frac{\alpha_1-\alpha_2}{2} \pi x) \bigg]  \nonumber \label{eqn:simplified_v}   \\
& \\
&  u(x,~N_0, s^*) \approx \frac{1}{x^2} cos(\frac{\alpha_1}{2} \pi x) + ( 1 - \frac{1}{x^2} ) cos( \frac{\alpha_2}{2} \pi x)  \label{eqn:simplified_u} \nonumber \\
\end{align}

\begin{figure}[ht!]
	\centering
	\includegraphics[trim= 200 0 220 0, clip, width=1\columnwidth]{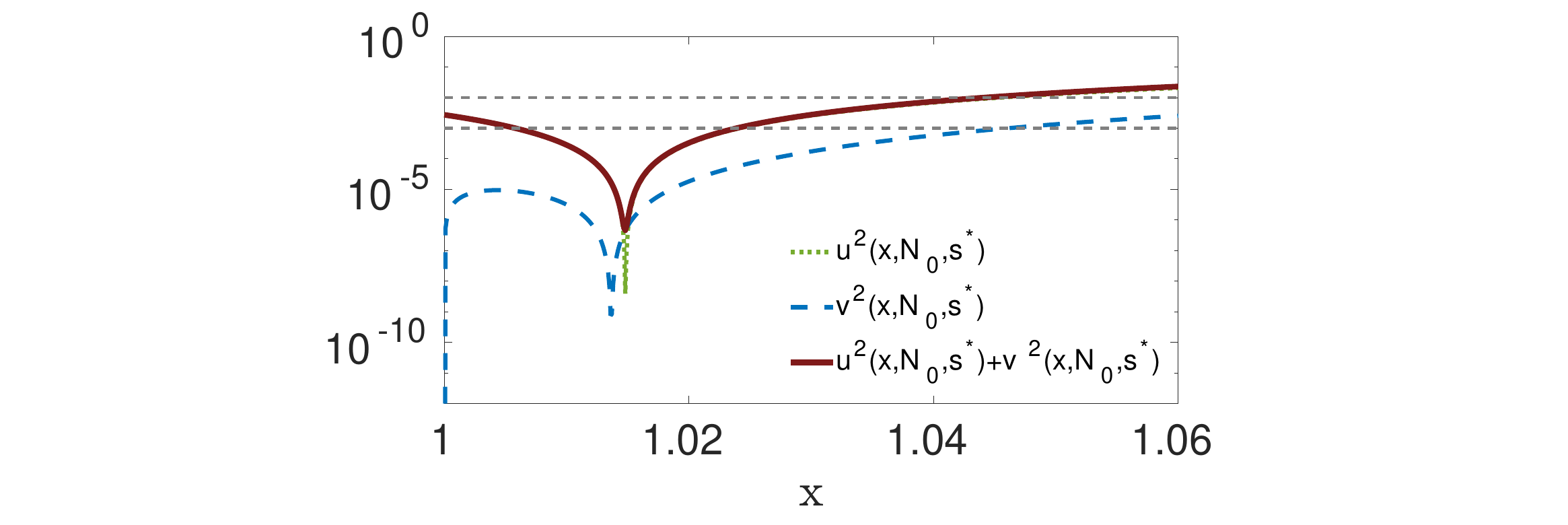}
	\caption{\label{fig:crossings_complex_trasmission} Illustration of $u^2(x, N_0, s^*)$ (Dotted black curve), $v^2(x, N_0, s^*)$ (Dashed black curve) and $u^2(x,  N_0, s^*) + v^2(x, N_0, s^*)$ (Solid  gray curve) with $N_0=30$ and $s^*=0.02$. The dashed gray horizontal lines represent $y=1 \times 10^{-2}$ (on top) and $y=10^{-3}$ (on bottom).}
\end{figure}

\begin{figure}[ht!]
	\centering
	\includegraphics[trim= 0 0 0 0, clip, width=1\columnwidth]{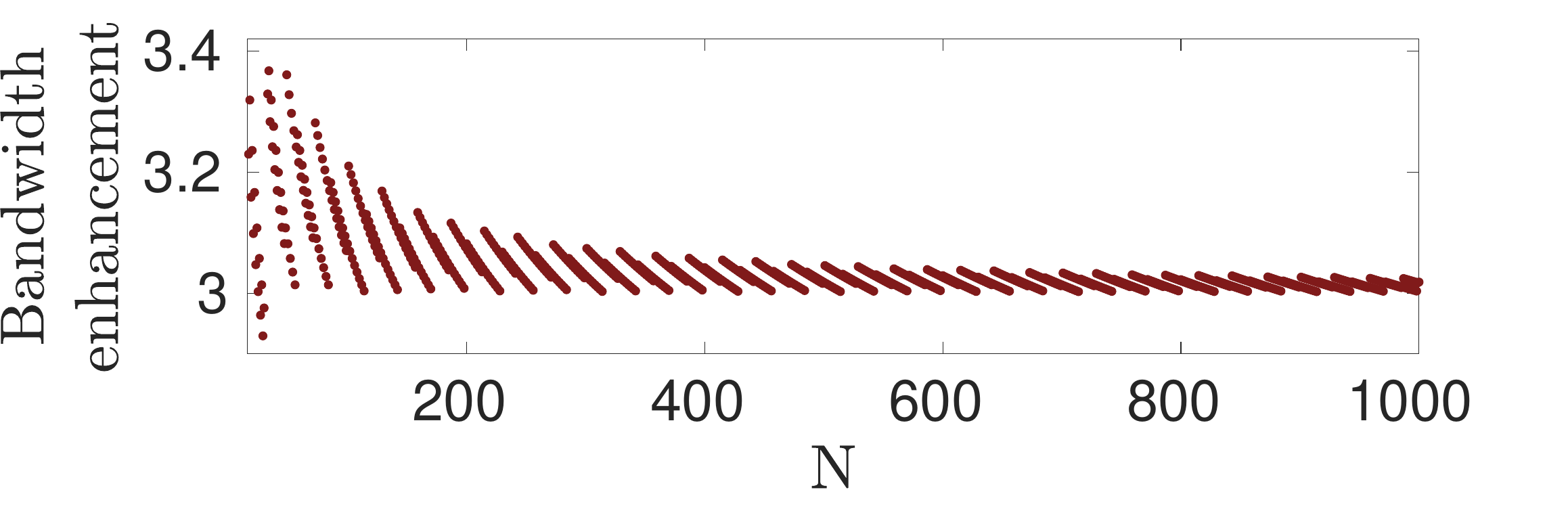}
	\caption{\label{fig:enhancement} Calculated bandwidth enhancement of a $\pi$-phase-shifted grating compared to a standard grating, given by $\Omega(N_0, s) = \sqrt{x(N_0,s)^2-1} / \sqrt{1.005^2-1}$. In particular, we find $\Omega(N_0, s)>3$ for all $N_0 \ge 32$.}
\end{figure}

For any given $N_0$ and $s^*$, $x_{0.01} (N_0, s^*)$ can be numerically calculated. Example of calculated $u^2(x, N_0, s^*)+v^2(x, N_0,s^*)$ for $N_0=30$ and $s^*=0.02$ is shown in figure \ref{fig:crossings_complex_trasmission}. Since $\phi(\lambda)/K \approx \sqrt{x^2(\lambda)-1}$, the relative enhancement of 99 \% conversion bandwidth compared to a standard grating with identical parameters can be estimated by the ratio $\Omega(N_0, s) = \frac{\sqrt{x_{\textcolor{black}{0.01}}(N_0,s)^2-1}}{\sqrt{1.005^2-1}}$. Computed $\Omega(N_0, s)$ for $16 \le N_0 \le \textcolor{black}{1000}$ and $s=0.02$ are presented in figure \ref{fig:enhancement}, showing at least 2.9 times relative bandwidth enhancement. Although smaller $N$ corresponds to larger absolute bandwidth, it is harder to control due to large jumps between spectra during grating writing. 

In order to predict the spectral shape above $99$ \% mode conversion, we notice that for any $16 \le N_0 \le 1000$, $-0.02 \le s^* \le 0.02$  and $1 \le x \le 1.045$, we have $v^2(x, N_0, s^*) < 10^{-3}$. Thus the spectral shape of $\vert T_{1,1} \vert^2$ is dominated by $u^2(x, N_0, s^*)$. To prove that $u(x, N_0, s^*)$ attains zero only once in $1 \le x \le 1.045$, let us check its sign at $x=1$ and $x=\frac{1}{\alpha_1}$. 

\begin{figure*}[ht!]
	\centering
	\includegraphics[trim= 10 0 5 0, clip, width=2\columnwidth]{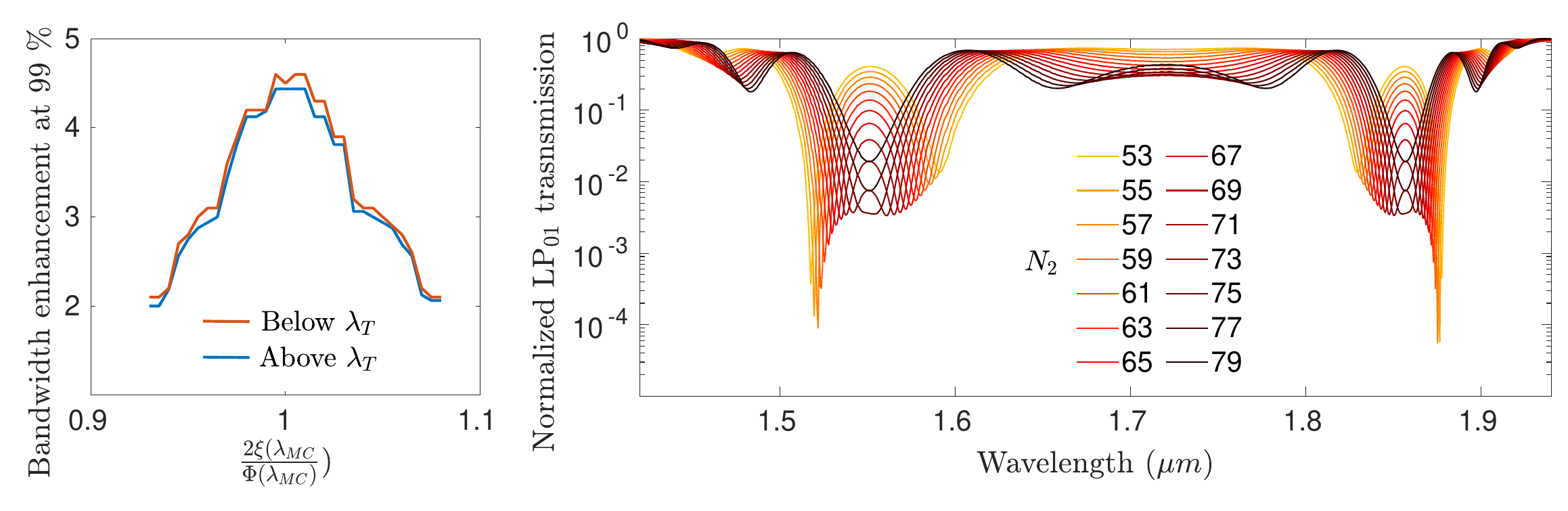}
	\caption{\label{fig:inexact} (a) Calculated LP$_{01}$ transmission spectra after LP$_{01}$-LP$_{02}$ mode conversion by a single-phase-shifted grating with inexact phase shift of $\xi(\lambda_{MC})=\pi/1.04$, plotted for different $N_2$. $N_1=15$ corresponding to maximum transmission for $N_0=30$ periods of a uniform grating with identical parameters. For the spectral dips at shorter wavelengths than the critical wavelength $\lambda_T$ given by $\frac{\partial \Phi(\lambda_T)}{\partial \lambda} = 0$, the dip with smaller wavelength grows and reverts earlier. For wavelengths larger than $\lambda_T$, the spectral dip with larger wavelength grows and reverts earlier. (b) For different values of $\frac{2\xi(\lambda_{MC})}{\Phi(\lambda_{MC})}$ and $N_1=15$, the best bandwidth enhancement calculated over different $N_2$.}
\end{figure*}

\begin{equation} \label{table:u(x)_signs}
\begin{array}{| c | c | c | c |}
\hline 
x & \frac{1}{x^2} cos(\alpha_{1} \frac{\pi}{2} x) & ( 1 - \frac{1}{x^2} ) cos( \alpha_{2} \frac{\pi}{2} x)& u(x, N_0) \\
\hline  \hline 
1 & cos(\alpha_{1} \frac{\pi}{2} ) >0 & 0 & >0 \\
\hline 
\frac{1}{\alpha_1} & 0 &  ( 1 - \alpha_1^2 ) cos( \frac{\pi}{2} \frac{\alpha_2}{\alpha_1})  <0 & <0 \\
\hline
\end{array} \nonumber
\end{equation}

Thus $u(x, N_0, s^*)=0$ for some $x_0(N_0, s) \in ( 1 , \frac{1}{\alpha_1} )$, which can be shown to be unique as follows. $\frac{1}{x^2} cos(\alpha_{1} \frac{\pi}{2} x)$ is monotonically decreasing in $1 \le x \le \frac{2}{\alpha_1}$. Wherever $f(x)=( 1 - \frac{1}{x^2} ) cos( \alpha_{2} \frac{\pi}{2} x)$ has an extremum, $f'(x)=(1 - \frac{1}{x^2})  \frac{\alpha_2 \pi}{2} sin( \frac{\alpha_2 \pi}{2} x) + \frac{2}{x^2} sin( \frac{\alpha_2 \pi}{2} x))=0$. Rearranging we get $cot(\alpha_2 \frac{\pi}{2} x)= -\frac{\alpha_2 \pi}{4} x (x^2-1)$, which can only hold for $x >1.2$ whenever $2.91 \le \alpha_2 \le 2.95$. Thus for $1 \le x \le 1.045$, $x_0(N_0, s)$ is the only possible root of $u(x, N_0, s^*)$. This proves that the proposed single-phase-shifted mode converter will have two spectral dips $\lambda_{-} < \lambda_{MC} < \lambda_{+}$ approximately satisfying $( 1 + \frac{\phi^2(\lambda_{\pm})}{K^2(\lambda_{\pm})} )^\frac{1}{2}  \approx x_0(N_0,s)$.

The 3-fold enhancement of 99 \% conversion bandwidth is a guaranteed estimate for any waveguide satisfying $s \le 0.02$. The actual enhancement can be larger, as we can see from simple simulation of LP$_{01}$-LP$_{02}$ mode conversion in a step-index fiber of radius 5.0 $\mu$m and 14 \% GeO$_2$ doping in the core. $\beta_k(\lambda)$ were calculated semi-analytically using V-b diagram \cite{okamoto_chapter_2006} and Sellmeier coefficients \cite{fleming_dispersion_1984}. Bandwidth enhancement calculated directly from spectra for $N_0=50$ is illustrated in figure \ref{fig:enhancement}a for different values of ${2 \xi(\lambda_{MC})}/{\Phi(\lambda_{MC})}$, which shows 3-fold enhancement in the range $0.97 \le {2 \xi(\lambda_{MC})}/{\Phi(\lambda_{MC})} \le 1.03$, peaking at $\xi(\lambda_{MC})=\pi$.  For ${2 \xi(\lambda_{MC})}/{\Phi(\lambda_{MC})}<0.93$ and ${2 \xi(\lambda_{MC})}/{\Phi(\lambda_{MC})}>1.08$ the normalized transmission intensity never goes below 0.01, thereby barring any enhancement. Degradation of bandwidth enhancement for inexact $\xi(\lambda_{MC})$ is directly illustrated by calculated spectra over $N_2$, as illustrated in \ref{fig:inexact}b for assumed $\Phi(\lambda)= 1.04 \Lambda_{MC} (\beta_1(\lambda) -\beta_2(\lambda) )$ and inexact $\xi(\lambda)= \pi / 1.04$. Below the critical wavelength ($\frac{\partial \Phi(\lambda)}{\partial \lambda}>0$), amongst the coalescing spectra dips on the two sides of $\lambda_{MC}$, the spectral dip with smaller $\lambda$ reaches minimum first with increasing $N_2$. The behavior switches above the critical wavelength. Whenever $\xi(\lambda) \ne \pi$, the minima of the spectral dips of $T_{1,1}(\lambda)$ on two sides of $\lambda_{MC}$ evolve at different rates with $N_2$, hampering flattening of the conversion spectrum. 

Equation \ref{eqn:single_gratings} does not capture the effect of duty cycle, which can be addressed by simulating gratings using "discretized-medium" approximation \cite{wang_layer_2001}. We simulated $\ge$ 99.9 \% conversion for duty cycles in the range $\big( 0.35, 0.65 \big)$ for the first 4 LP modes \cite{gloge}. Thus our solution is robust to duty cycle. Other than the critical accuracy of the $\pi$-phase-shift, the physical non-uniformity of actual waveguides over lengthscales of few grating periods can also affect the phase-interplay and degrade performance. Shorter gratings are more robust to such variations.  

In conclusion, we prove at least $2.9$ times bandwidth enhancement at above $99$ \% mode conversion by a $\pi$-phase-shifted grating which has minimum number of periods $N_1$ and $N_2$ in first and second segment such that the transmission intensity of the original mode at the resonance wavelength is $\le 50 \%$ after writing only $N_1$ periods and $<10^{-2}$ after writing both segments. This deterministic solution makes broadband mode converter gratings independent of material parameters, mode-pair and even grating-writing method, thereby relaxing major experimental bottlenecks.

\section{Disclosures}
The author declares no conflicts of interest. 

\bibliography{Draft_corr}

%merlin.mbs apsrev4-1.bst 2010-07-25 4.21a (PWD, AO, DPC) hacked
%Control: key (0)
%Control: author (8) initials jnrlst
%Control: editor formatted (1) identically to author
%Control: production of article title (-1) disabled
%Control: page (0) single
%Control: year (1) truncated
%Control: production of eprint (0) enabled
\begin{thebibliography}{16}%
\makeatletter
\providecommand \@ifxundefined [1]{%
 \@ifx{#1\undefined}
}%
\providecommand \@ifnum [1]{%
 \ifnum #1\expandafter \@firstoftwo
 \else \expandafter \@secondoftwo
 \fi
}%
\providecommand \@ifx [1]{%
 \ifx #1\expandafter \@firstoftwo
 \else \expandafter \@secondoftwo
 \fi
}%
\providecommand \natexlab [1]{#1}%
\providecommand \enquote  [1]{``#1''}%
\providecommand \bibnamefont  [1]{#1}%
\providecommand \bibfnamefont [1]{#1}%
\providecommand \citenamefont [1]{#1}%
\providecommand \href@noop [0]{\@secondoftwo}%
\providecommand \href [0]{\begingroup \@sanitize@url \@href}%
\providecommand \@href[1]{\@@startlink{#1}\@@href}%
\providecommand \@@href[1]{\endgroup#1\@@endlink}%
\providecommand \@sanitize@url [0]{\catcode `\\12\catcode `\$12\catcode
  `\&12\catcode `\#12\catcode `\^12\catcode `\_12\catcode `\%12\relax}%
\providecommand \@@startlink[1]{}%
\providecommand \@@endlink[0]{}%
\providecommand \url  [0]{\begingroup\@sanitize@url \@url }%
\providecommand \@url [1]{\endgroup\@href {#1}{\urlprefix }}%
\providecommand \urlprefix  [0]{URL }%
\providecommand \Eprint [0]{\href }%
\providecommand \doibase [0]{http://dx.doi.org/}%
\providecommand \selectlanguage [0]{\@gobble}%
\providecommand \bibinfo  [0]{\@secondoftwo}%
\providecommand \bibfield  [0]{\@secondoftwo}%
\providecommand \translation [1]{[#1]}%
\providecommand \BibitemOpen [0]{}%
\providecommand \bibitemStop [0]{}%
\providecommand \bibitemNoStop [0]{.\EOS\space}%
\providecommand \EOS [0]{\spacefactor3000\relax}%
\providecommand \BibitemShut  [1]{\csname bibitem#1\endcsname}%
\let\auto@bib@innerbib\@empty
%</preamble>
\bibitem [{\citenamefont {Gloge}(1971)}]{gloge}%
  \BibitemOpen
  \bibfield  {author} {\bibinfo {author} {\bibfnamefont {D.}~\bibnamefont
  {Gloge}},\ }\href {\doibase 10.1364/AO.10.002252} {\bibfield  {journal}
  {\bibinfo  {journal} {Applied Optics}\ }\textbf {\bibinfo {volume} {10}},\
  \bibinfo {pages} {2252} (\bibinfo {year} {1971})}\BibitemShut {NoStop}%
\bibitem [{\citenamefont {Poole}\ \emph {et~al.}(1994)\citenamefont {Poole},
  \citenamefont {Wiesenfeld}, \citenamefont {DiGiovanni},\ and\ \citenamefont
  {Vengsarkar}}]{poole_optical_1994}%
  \BibitemOpen
  \bibfield  {author} {\bibinfo {author} {\bibfnamefont {C.~D.}\ \bibnamefont
  {Poole}}, \bibinfo {author} {\bibfnamefont {J.~M.}\ \bibnamefont
  {Wiesenfeld}}, \bibinfo {author} {\bibfnamefont {D.~J.}\ \bibnamefont
  {DiGiovanni}}, \ and\ \bibinfo {author} {\bibfnamefont {A.~M.}\ \bibnamefont
  {Vengsarkar}},\ }\href {\doibase 10.1109/50.337486} {\bibfield  {journal}
  {\bibinfo  {journal} {Journal of Lightwave Technology}\ }\textbf {\bibinfo
  {volume} {12}},\ \bibinfo {pages} {1746} (\bibinfo {year}
  {1994})}\BibitemShut {NoStop}%
\bibitem [{\citenamefont {Namiki}\ and\ \citenamefont
  {Haus}(1997)}]{namiki_noise_1997}%
  \BibitemOpen
  \bibfield  {author} {\bibinfo {author} {\bibfnamefont {S.}~\bibnamefont
  {Namiki}}\ and\ \bibinfo {author} {\bibfnamefont {H.}~\bibnamefont {Haus}},\
  }\href {\doibase 10.1109/3.572138} {\bibfield  {journal} {\bibinfo  {journal}
  {IEEE Journal of Quantum Electronics}\ }\textbf {\bibinfo {volume} {33}},\
  \bibinfo {pages} {649} (\bibinfo {year} {1997})}\BibitemShut {NoStop}%
\bibitem [{\citenamefont {Ramachandran}\ \emph {et~al.}(2002)\citenamefont
  {Ramachandran}, \citenamefont {Wang},\ and\ \citenamefont
  {Yan}}]{ramachandran_bandwidth_2002}%
  \BibitemOpen
  \bibfield  {author} {\bibinfo {author} {\bibfnamefont {S.}~\bibnamefont
  {Ramachandran}}, \bibinfo {author} {\bibfnamefont {Z.}~\bibnamefont {Wang}},
  \ and\ \bibinfo {author} {\bibfnamefont {M.}~\bibnamefont {Yan}},\ }\href
  {\doibase 10.1364/OL.27.000698} {\bibfield  {journal} {\bibinfo  {journal}
  {Optics Letters}\ }\textbf {\bibinfo {volume} {27}},\ \bibinfo {pages} {698}
  (\bibinfo {year} {2002})}\BibitemShut {NoStop}%
\bibitem [{\citenamefont {Zheng}\ \emph {et~al.}(2016)\citenamefont {Zheng},
  \citenamefont {Li}, \citenamefont {Li},\ and\ \citenamefont
  {Wu}}]{zheng_all_2016}%
  \BibitemOpen
  \bibfield  {author} {\bibinfo {author} {\bibfnamefont {Y.}~\bibnamefont
  {Zheng}}, \bibinfo {author} {\bibfnamefont {Y.}~\bibnamefont {Li}}, \bibinfo
  {author} {\bibfnamefont {W.}~\bibnamefont {Li}}, \ and\ \bibinfo {author}
  {\bibfnamefont {J.}~\bibnamefont {Wu}},\ }in\ \href {\doibase
  10.1364/CLEO_AT.2016.JTu5A.108} {\emph {\bibinfo {booktitle} {Conference on
  {Lasers} and {Electro}-{Optics} (2016), paper {JTu5A}.108}}}\ (\bibinfo
  {publisher} {Optical Society of America},\ \bibinfo {year} {2016})\ p.\
  \bibinfo {pages} {JTu5A.108}\BibitemShut {NoStop}%
\bibitem [{\citenamefont {Savolainen}\ \emph {et~al.}(2014)\citenamefont
  {Savolainen}, \citenamefont {Kristensen}, \citenamefont {Grüner-Nielsen},\
  and\ \citenamefont {Balling}}]{savolainen_broadband_2014}%
  \BibitemOpen
  \bibfield  {author} {\bibinfo {author} {\bibfnamefont {J.~M.}\ \bibnamefont
  {Savolainen}}, \bibinfo {author} {\bibfnamefont {P.}~\bibnamefont
  {Kristensen}}, \bibinfo {author} {\bibfnamefont {L.}~\bibnamefont
  {Grüner-Nielsen}}, \ and\ \bibinfo {author} {\bibfnamefont {P.}~\bibnamefont
  {Balling}},\ }\href {\doibase 10.1109/LPT.2014.2326884} {\bibfield  {journal}
  {\bibinfo  {journal} {IEEE Photonics Technology Letters}\ }\textbf {\bibinfo
  {volume} {26}},\ \bibinfo {pages} {1454} (\bibinfo {year}
  {2014})}\BibitemShut {NoStop}%
\bibitem [{\citenamefont {Jung}\ \emph {et~al.}(2015)\citenamefont {Jung},
  \citenamefont {Alam},\ and\ \citenamefont {Richardson}}]{jung_compact_2015}%
  \BibitemOpen
  \bibfield  {author} {\bibinfo {author} {\bibfnamefont {Y.}~\bibnamefont
  {Jung}}, \bibinfo {author} {\bibfnamefont {S.~U.}\ \bibnamefont {Alam}}, \
  and\ \bibinfo {author} {\bibfnamefont {D.~J.}\ \bibnamefont {Richardson}},\
  }in\ \href {\doibase 10.1109/ECOC.2015.7341644} {\emph {\bibinfo {booktitle}
  {2015 {European} {Conference} on {Optical} {Communication} ({ECOC})}}}\
  (\bibinfo {year} {2015})\ pp.\ \bibinfo {pages} {1--3}\BibitemShut {NoStop}%
\bibitem [{\citenamefont {Chan}\ and\ \citenamefont
  {Chiang}(2005)}]{chan_analysis_2005}%
  \BibitemOpen
  \bibfield  {author} {\bibinfo {author} {\bibfnamefont {F.~Y.~M.}\
  \bibnamefont {Chan}}\ and\ \bibinfo {author} {\bibfnamefont {K.~S.}\
  \bibnamefont {Chiang}},\ }\href {\doibase 10.1016/j.optcom.2004.09.078}
  {\bibfield  {journal} {\bibinfo  {journal} {Optics Communications}\ }\textbf
  {\bibinfo {volume} {244}},\ \bibinfo {pages} {233} (\bibinfo {year}
  {2005})}\BibitemShut {NoStop}%
\bibitem [{\citenamefont {Wang}\ \emph {et~al.}(2017)\citenamefont {Wang},
  \citenamefont {Wu}, \citenamefont {Chen}, \citenamefont {Jin},\ and\
  \citenamefont {Chiang}}]{wang_ultra-broadband_2017}%
  \BibitemOpen
  \bibfield  {author} {\bibinfo {author} {\bibfnamefont {W.}~\bibnamefont
  {Wang}}, \bibinfo {author} {\bibfnamefont {J.}~\bibnamefont {Wu}}, \bibinfo
  {author} {\bibfnamefont {K.}~\bibnamefont {Chen}}, \bibinfo {author}
  {\bibfnamefont {W.}~\bibnamefont {Jin}}, \ and\ \bibinfo {author}
  {\bibfnamefont {K.~S.}\ \bibnamefont {Chiang}},\ }\href {\doibase
  10.1364/OE.25.014341} {\bibfield  {journal} {\bibinfo  {journal} {Optics
  Express}\ }\textbf {\bibinfo {volume} {25}},\ \bibinfo {pages} {14341}
  (\bibinfo {year} {2017})}\BibitemShut {NoStop}%
\bibitem [{\citenamefont {Liu}\ \emph {et~al.}(2020)\citenamefont {Liu},
  \citenamefont {Liu}, \citenamefont {Zhao},\ and\ \citenamefont
  {Mou}}]{liu_bandwidth_2020}%
  \BibitemOpen
  \bibfield  {author} {\bibinfo {author} {\bibfnamefont {Z.}~\bibnamefont
  {Liu}}, \bibinfo {author} {\bibfnamefont {Y.}~\bibnamefont {Liu}}, \bibinfo
  {author} {\bibfnamefont {X.}~\bibnamefont {Zhao}}, \ and\ \bibinfo {author}
  {\bibfnamefont {C.}~\bibnamefont {Mou}},\ }\href {\doibase 10.1364/OE.408623}
  {\bibfield  {journal} {\bibinfo  {journal} {Optics Express}\ }\textbf
  {\bibinfo {volume} {28}},\ \bibinfo {pages} {31882} (\bibinfo {year}
  {2020})},\ \bibinfo {note} {publisher: Optical Society of
  America}\BibitemShut {NoStop}%
\bibitem [{\citenamefont {Abrishamian}\ and\ \citenamefont
  {Morishita}(2015)}]{abrishamian_cascade_2015}%
  \BibitemOpen
  \bibfield  {author} {\bibinfo {author} {\bibfnamefont {F.}~\bibnamefont
  {Abrishamian}}\ and\ \bibinfo {author} {\bibfnamefont {K.}~\bibnamefont
  {Morishita}},\ }\href {\doibase 10.1587/transele.E98.C.512} {\bibfield
  {journal} {\bibinfo  {journal} {IEICE Transactions on Electronics}\ }\textbf
  {\bibinfo {volume} {E98.C}},\ \bibinfo {pages} {512} (\bibinfo {year}
  {2015})}\BibitemShut {NoStop}%
\bibitem [{\citenamefont {Alcusa-Sáez}\ \emph {et~al.}(2016)\citenamefont
  {Alcusa-Sáez}, \citenamefont {Díez},\ and\ \citenamefont
  {Andrés}}]{alcusa-2016}%
  \BibitemOpen
  \bibfield  {author} {\bibinfo {author} {\bibfnamefont {E.}~\bibnamefont
  {Alcusa-Sáez}}, \bibinfo {author} {\bibfnamefont {A.}~\bibnamefont {Díez}},
  \ and\ \bibinfo {author} {\bibfnamefont {M.~V.}\ \bibnamefont {Andrés}},\
  }\href {\doibase 10.1364/OE.24.004899} {\bibfield  {journal} {\bibinfo
  {journal} {Optics Express}\ }\textbf {\bibinfo {volume} {24}},\ \bibinfo
  {pages} {4899} (\bibinfo {year} {2016})}\BibitemShut {NoStop}%
\bibitem [{\citenamefont {Basu}(2020)}]{basu}%
  \BibitemOpen
  \bibfield  {author} {\bibinfo {author} {\bibfnamefont {S.}~\bibnamefont
  {Basu}},\ }\href {http://ol.osa.org/abstract.cfm?URI=ol-45-19-5518}
  {\bibfield  {journal} {\bibinfo  {journal} {Opt. Lett.}\ }\textbf {\bibinfo
  {volume} {45}},\ \bibinfo {pages} {5518} (\bibinfo {year}
  {2020})}\BibitemShut {NoStop}%
\bibitem [{\citenamefont {Wang}\ and\ \citenamefont
  {Erdogan}(2001)}]{wang_layer_2001}%
  \BibitemOpen
  \bibfield  {author} {\bibinfo {author} {\bibfnamefont {L.}~\bibnamefont
  {Wang}}\ and\ \bibinfo {author} {\bibfnamefont {T.}~\bibnamefont {Erdogan}},\
  }\href {\doibase 10.1049/el:20010142} {\bibfield  {journal} {\bibinfo
  {journal} {Electronics Letters}\ }\textbf {\bibinfo {volume} {37}},\ \bibinfo
  {pages} {154} (\bibinfo {year} {2001})}\BibitemShut {NoStop}%
\bibitem [{\citenamefont {Okamoto}(2006)}]{okamoto_chapter_2006}%
  \BibitemOpen
  \bibfield  {author} {\bibinfo {author} {\bibfnamefont {K.}~\bibnamefont
  {Okamoto}},\ }in\ \href {\doibase 10.1016/B978-012525096-2/50004-0} {\emph
  {\bibinfo {booktitle} {Fundamentals of {Optical} {Waveguides} ({Second}
  {Edition})}}}\ (\bibinfo  {publisher} {Academic Press},\ \bibinfo {address}
  {Burlington},\ \bibinfo {year} {2006})\ pp.\ \bibinfo {pages}
  {57--158}\BibitemShut {NoStop}%
\bibitem [{\citenamefont {Fleming}(1984)}]{fleming_dispersion_1984}%
  \BibitemOpen
  \bibfield  {author} {\bibinfo {author} {\bibfnamefont {J.~W.}\ \bibnamefont
  {Fleming}},\ }\href {\doibase 10.1364/AO.23.004486} {\bibfield  {journal}
  {\bibinfo  {journal} {Applied Optics}\ }\textbf {\bibinfo {volume} {23}},\
  \bibinfo {pages} {4486} (\bibinfo {year} {1984})}\BibitemShut {NoStop}%
\end{thebibliography}%

\end{document}